\documentclass[12pt]{article}
\usepackage{amsfonts}
\usepackage{amsmath}
\usepackage{amssymb}
\newcommand{\be}{\begin{equation}}
\newcommand{\ba}{\begin{eqnarray}}
\newcommand{\ee}{\end{equation}}
\newcommand{\ea}{\end{eqnarray}}

\DeclareMathOperator{\sech}{sech}

\begin{document}

\title{Scattering amplitudes for the rationally extended $PT$ symmetric complex potentials}

\author{Nisha Kumari$^{a}$\footnote{e-mail address: nishaism0086@gmail.com (N.K)}, Rajesh Kumar Yadav$^{b}$\footnote {e-mail address: rajeshastrophysics@gmail.com (R.K.Y)}, Avinash Khare$^{c}$\footnote{e-mail: khare@physics.unipune.ac.in (A.K)}, Bijan Bagchi$^{d}$\footnote{e-mail address:bbagchi123@gmail.com (B.B) } and \\ Bhabani Prasad Mandal$^{a}$\footnote{e-mail address: bhabani.mandal@gmail.com (B.P.M).}}
 \maketitle
 {$~^a$ Department of Physics, Banaras Hindu University, Varanasi-221005, India.\\
 $~^b$ Department of Physics, S. P. College, S. K. M. University, Dumka-814101, India.\\
 $~^c$ Department of Physics, Savitribai Phule Pune University, Pune-411007, India.\\
$~^d$  Department of Physics, School of Natural Science, Shiv Nadar University, Greater Noida, UP-201314 India.
}

\begin{abstract}
In this paper, we consider the rational extensions of two different $PT$ symmetric complex potentials namely the asymptotically
vanishing   Scarf II and asymptotically non-vanishing  Rosen-Morse II [ RM-II]
potentials and obtain  bound state eigenfunctions in terms of newly found
exceptional $X_m$ Jacobi polynomials and also some new type of orthogonal
polynomials respectively. By considering the asymptotic behaviour of the
exceptional polynomials, we obtain the reflection and transmission
amplitudes for them and discuss the various novel properties of
the corresponding amplitudes.

\end{abstract}

\section{Introduction}

Recent discovery of exceptional orthogonal polynomials (EOPs) (also known as $X_m$ Laguerre and $X_m$ Jacobi
orthogonal polynomials with $m=0,1,2,...,.$) \cite{dnr1,dnr2,xm1,xm2}, has
motivated researchers to search for new exactly solvable potentials whose
bound state
eigenfunctions are in terms of these EOPs. Interestingly it has been observed
that the newly discovered potentials are the rational extension
of the corresponding conventional potentials \cite{que,bqr,os,dim}.
These potentials are rationally extended trigonometric Scarf (Scarf-I),
generalized P\"oschl Teller (GPT) and radial oscillator
 potentials. Remarkably, these potentials can be cast in the framework of
supersymmetric quantum mechanics (SQM) and they have been shown to satisfy
the translational shape invariance (SI) condition \cite{cks}.

In the last few years, researchers have also discovered another category of
rationally extended potentials
whose bound state eigenfunctions are not in terms of the EOPs but rather
they are written in terms of some new polynomials (which in turn
can be expressed in terms of the classical orthogonal polynomials). These
rationally extended potentials do not satisfy the usual
translational SI property, but instead they satisfy an unfamiliar type of
extended SI property. The potentials belong to this category are
rationally extended Rosen-Morse II, Eckart etc \cite{op4,op5}. It is important
 here to note that the bound state eigenvalues of all these rationally
extended potentials are the same as those of their conventional counterparts i.e. they are iso-spectral.

Recently, fully consistent quantum theories have been developed for certain class of non-hermitian systems. The parity (P)
and time reversal (T), two important discrete symmetries play extremely important role in such formulations. It has been shown
that the $PT$ symmetric (combined parity and time reversal symmetry)
non-hermitian systems can have the  spectrum real if the
$PT$ symmetric is unbroken \cite{bender} and consequently a fully consistent
quantum theory with unitary time evolution is constructed in a modified
Hilbert space \cite{ali}. Since the last one and half decades, the $PT$
symmetric non-hermitian systems have been developed considerably and have
found several applications in optics as well as other branches of physics
\cite {mus,rutter,makris, pati,bender2,pt4,pt5,pt6}. Some of the important properties such as the existence of spectral singularity
(divergence of both reflection and transmission co-efficients at a point)
\cite{ss1,ss2,ss3,ss4,ss5}, exceptional points (coalesce of two energy
levels along with their eigenfunctions at certain points in the complex plane)
 \cite{ep1,ep2,ep3,ep4}, Handedness effect (dependence of
reflection coefficient on the direction of wave) \cite{zafer2} have been
also observed in such systems.

In the last few years while much work has been done on the
PT-symmetric potentials, we find that comparatively not enough work has
been done about the rationally extended $PT$ symmetric (REPTS) complex potentials.
However, considerable work has already been done about the rationally extended
real (Hermitian) potentials and bound state eigenstates as well as scattering
matrix has already been reported for them \cite{scatt1,scatt2,scatt3,scatt4}.
For example, while rational extension of real (Hermitian) Rosen-Morse II
potential has been extensively discussed \cite{op4}, it seems
there has been no discussion about the corresponding rationally extended,
complex $PT$ invariant Rosen-Morse II potential.
Further, to the best of our knowledge, only one REPTS complex potential corresponding to the conventional
$PT$ symmetric complex Scarf II potential \cite{bqpt10} has so far been
discussed in the literature.
In particular, it has been shown that the bound state eigenfunctions of
rationally extended PT-symmetric complex Scarf II potential are in terms
of $X_m$ Jacobi EOPs \cite{bb12}. However, as far as we know,
 the corresponding reflection and transmission amplitudes $r(k)$ and $t(k)$
respectively are not known in the literature.

The purpose of this paper is to fill the missing gap and discuss bound state
eigenstates as well as reflection and transmission coefficients of few
rationally extended PT-symmetric complex potentials. In particular,
in this paper, we obtain bound state eigenfunctions as well as $r(k)$ and
$t(k)$ for the rationally extended $PT$ symmetric complex Rosen-Morse II
potential. Further, we construct one more complex and $PT$ symmetric
rationally
extended Rosen-Morse II potential and obtain its bound state eigenfunctions.
Further, we obtain $r(k)$ and $t(k)$ for both the forms of complex
PT-symmetric rationally extended Rosen-Morse II potentials.
Recently an important symmetry i.e., the parametric symmetry in the $PT$
symmetric conventional as well as rationally extended Scarf II potentials has
been observed. Under this symmetry while the $PT$ symmetric Scarf II potential
 remains invariant (and has two different sets of bound states), however
the rationally extended $PT$ symmetric complex Scarf II potential \cite{para}
is not invariant and one gets another rationally extended $PT$ symmetric
complex Scarf II potential. It is thus of interest to inquire how $r(k)$ and
$t(k)$ change under the parametric symmetry.

 The organization of this paper is as follows:

In section $2$, we briefly discuss the bound state eigenfunctions of the
rationally extended $PT$ symmetric complex Scarf II and Rosen-Morse
II potentials. The parametric symmetry in the case of the extended $PT$
symmetric complex Scarf II is also discussed. The scattering amplitude,
i.e. $r(k)$ and $t(k)$ for these potentials is obtained in section $3$ where
some interesting properties of $r(k)$ and $t(k)$ are highlighted.
Finally we summarize the work done in this paper in section $4$.


\section{Bound states: REPTS potentials}

In this section, we briefly discuss the bound state eigenfunctions of the
two rationally extended complex $PT$ symmetric potentials .

\subsection{REPTS complex Scarf-II potential}

The well known complex and $PT$ symmetric conventional Scarf II potential  \cite{pt1} which is translationally SI is
given by
\be\label{scarfpot}
V(x,A,B)=(-B^2-A(A+1))\sech^2 x+iB(2A+1)\sech x\tanh x;\quad A > 0\,.
\ee
The bound states energy eigenvalues and the eigenfunctions are known to be\cite{bb12}
\be\label{scarfev}
E^{(A)}_n=-(A-n)^2;\qquad n=0,1,2....n_{\mbox{max}}<A,
\ee
and
\be\label{scarfef}
\psi^{(A,B)}_n (x)=N^{(\alpha,\beta )}_n (\sech x)^A \exp (-iB \tan^{-1}(\sinh x))P^{(\alpha,\beta)}_n(i\sinh x);\quad -\infty < x < \infty,
\ee
where $P^{(\alpha,\beta)}_n(z)$ is classical Jacobi polynomial,
$N^{(\alpha,\beta )}_n$ is the normalization
constant and the parameters $\alpha=B-A-\frac{1}{2}$ and $\beta=-B-A-\frac{1}{2}$.
An interesting symmetry of this conventional $PT$ symmetric complex Scarf II
potential has been observed
under a parametric transformation \cite{pt3} i.e.
$B\longleftrightarrow (A+\frac{1}{2})$. Under this transformation
 the potential given in Eq. (\ref{scarfpot}) remains invariant. This symmetry
implies another set of bound state solutions with the eigenstates given by
 \be\label{scarfev1}
E^{(B)}_n=-(B-n-\frac{1}{2})^2;\qquad n=0,1,2,...,n_{\mbox{max}}<B-\frac{1}{2},
\ee
and
\be\label{scarfwf1}
\psi^{(B\leftrightarrow A+\frac{1}{2})}_n(x)=N^{(\gamma,\delta)}_n (\sech x)^{B-\frac{1}{2}} \exp (-i(A+\frac{1}{2}) \tan^{-1}(\sinh x))P^{(\gamma,\delta)}_n(i\sinh x),
\ee
 where the new parameters $\gamma=-\alpha=A-B+\frac{1}{2}$ and $\delta=\beta=-A-B-\frac{1}{2}$.

At this point we observe that the potential given by Eq. (\ref{scarfpot})
has been extended rationally \cite {bb12} for any positive integer values of
$m$, i.e.
\ba\label{scarfextdpot}
V_{m}(x,A,B)&=&V(x,A,B)+2m(2B-m+1)+(2B-m+1)\nonumber\\
&\times &[(-2A-1)+(2B+1)i\sinh x]\frac{P^{(-\alpha,\beta)}_{m-1}(i\sinh x)}{P^{(-\alpha-1,\beta-1)}_{m}(i\sinh x)}\nonumber\\
&-&\frac{(2B-m+1)^2\cosh^2x}{2}\bigg (\frac{P^{(-\alpha,\beta )}_{m-1}(i\sinh x)}
{P^{(-\alpha-1,\beta-1 )}_{m}(i\sinh x)}\bigg )^2, \nonumber\\
\ea
where $V(x,A,B)$ is the conventional $PT$ symmetric complex Scarf II
potential given in Eq. (\ref{scarfpot}). Like the complex $PT$ symmetric
Scarf II potential given by Eq. (\ref{scarfpot}), these extended
potentials  are also SI under the translation of the parameters
$A \rightarrow A-1$ and the energy eigenvalues are (real)
and the same (i.e. isospectral) as that of the conventional one as
given by (\ref{scarfev}). The bound state
 eigenfunctions corresponding to the extended potential (\ref{scarfextdpot})
are given by
\be\label{scarfextdwf}
\psi^{(A,B)}_{n,m}(x)= N^{(\alpha,\beta )}_{n ,m}\times \frac{(1+\sinh^2 x)^{-\frac{A}{2}} \exp \left\{{-iB\tan^{-1}(\sinh x)}\right\}}{P^{(-\alpha-1 ,\beta-1 )}_{m} (i\sinh x)}
\hat{P}^{(\alpha ,\beta )}_{n +m} (i\sinh x),
\ee
where   $N^{(\alpha,\beta)}_{n ,m}$ is a normalization constant and
$\hat{P}^{(\alpha ,\beta )}_{n +m} (i\sinh x)$ is  $X_m$ Jacobi EOPs.
orthogonal polynomials. As expected, for $m=0$, the above rationally extended
potential (as well as the corresponding eigenfunctions) reduce to the
conventional one as given by Eqs. (\ref{scarfpot}) and (\ref{scarfev})
while for $m=1$ the potential corresponds to the case of $X_1$ EOPs
\cite{bqpt10}. Here it is important to note that unlike the conventional
potential Eq. (\ref{scarfpot}), the rationally extended potential
Eq. (\ref{scarfextdpot}) is not invariant under the transformation
$B\longleftrightarrow A+\frac{1}{2}$, rather it goes over to another extended
potential given by
\ba\label{extdscarf1}
V_m(x,B\leftrightarrow A+\frac{1}{2})&=&V(x)+2m(2A-m+2)+(2A-m+2)\nonumber\\
&\times &[(-2B)+(2A+2)i\sinh x]\frac{P^{(-\gamma,\delta)}_{m-1}(i\sinh x)}{P^{(-\gamma-1,\delta-1)}_{m}(i\sinh x)}\nonumber\\
&-&\frac{(2A-m+2)^2\cosh^2x}{2}\bigg (\frac{P^{(-\gamma,\delta )}_{m-1}(i\sinh x)}
{P^{(-\gamma-1,\delta-1 )}_{m}(i\sinh x)}\bigg )^2. \nonumber\\
\ea
The energy eigenvalues of this potential are isospectral to that of the
conventional potential obtained after the transformation
$B\longleftrightarrow A+\frac{1}{2}$ and are given by Eq. (\ref{scarfev1}).
The bound state eigenfunctons corresponding to this new
rationally extended potential are given by
\be\label{extdscarfwf1}
\psi^{(B\leftrightarrow A+\frac{1}{2})}_{(n,m)}(x) = N^{(\gamma,\delta)}_{(n,m)}\times \frac{(\sech x)^{A}\exp (-iB\tan^{-1}(\sinh x))}{P_{m}^{(-\gamma-1,\delta-1)}(i\sinh x)}\hat{P}_{n+m}^{(\gamma,\delta)}(i\sinh x).
\ee
The above rationally extended potential is also SI under the translation of parameter $B\longrightarrow B-1$ .

\subsection{REPTS complex RM-II potential}

This potential belongs to the second category of the rationally extended potentials whose bound state solutions are
not in the exact form of EOPs, they are in the form of some types of new polynomials \cite{op4}. These polynomials can be
written in terms of classical Jacobi polynomials. Further, these potentials
do not satisfy the usual SI property, but instead they satisfy an unusual
type of enlarge SI property where both the potential parameters
as well as the order of the polynomials change.

Using first-order SUSY QM, Quesne \cite {op4} obtained the rationally
extended real Rosen-Morse II (RM-II) potential by considering the
conventional real Rosen-Morse II potential. We can then obtain the rationally
extended $PT$ symmetric complex RM-II potentials by essentially following
the approach of \cite{op4} by considering the $PT$ symmetric complex RM-II
potential \cite{lev} (which can
be obtained by changing the potential parameter $B\longrightarrow iB$ in the conventional real potential) as
 \be\label{rm2pot}
V_{A,iB}(x)=-A(A+1){\rm sech^2} x +2iB\tanh x , \quad -\infty < x <\infty,
\ee
where $A >0$. The bound state energy eigenvalues turn out to be real
\be\label{rm2ev}
E^{(A,iB)}_n=-(A-n )^2+\frac{B^2}{(A-n)^2}, \qquad n=0,1,2....,n_{max}<A.
\ee
The corresponding wavefunctions in terms of classical Jacobi polynomials $P^{(A-n +\frac{iB}{A-n },A-n -\frac{iB}{A-n })}(z)$ are
\be\label{rm2wf}
\psi_{n }^{(A,iB)}(x)\propto (1-\tanh x)^{\frac{1}{2}(A-n +\frac{iB}{A-n})}(1+\tanh x)^{\frac{1}{2}(A-n -\frac{iB}{A-n })}
P_n ^{(A-n +\frac{iB}{A-n },A-n -\frac{iB}{A-n })}(z),
\ee
where $z=\tanh x$. Similar to the real case \cite {op4}, this complex $PT$
symmetric potential can be extended
rationally by determining all possible polynomials-type, nodeless solutions $\phi (x)$ of the Schr\"odinger equation
\be
-\frac{d^2\phi (x)}{dx^2}+V_{A, iB}(x)\phi (x)=E\phi (x)
\ee
with the factorization energy $E<E_{0}^{(A,iB)}=-A^2+\frac{B^2}{A^2}$.
Out of all the possible solutions of $\phi (x)$, two independent polynomial type solutions $\phi _{1}(x)$ and $\phi _{2}(x)$
with the energy $E_{1}$ and $E_{2}$ respectively have been constructed (for detail see Ref. \cite{op4}).
On putting some restrictions on the parameters $A$ and $B$, three acceptable polynomial-type,
nodeless solutions are obtained. By considering the
conventional potential with some different $A'$ i.e.
\be\label{rm2plus}
V^{(+)}(x)=V_{(A^{'},iB)}(x),
\ee
the rationally-extended $PT$ symmetric Rosen Morse-II potential $V^{(-)}(x)$ with given $A$ and $B$ is obtained as
 \ba\label{ratrm2}
 V^{(-)}(x)=V_{(A,iB,ext)}(x)= V_{(A,iB)}(x)+V_{(A,iB,rat)}(x),
 \ea
 where
 \be\label{extdrm2pot}
 V_{(A,iB,rat)}(x)=2(1-z^2)\left\{2z\frac{\dot{g}_m^{(A,iB)} }{g_m^{(A,iB)}}-(1-z^2)\Bigg[\frac{\ddot{g}_m^{(A,iB)} }{g_m^{(A,iB)}}-
\left(\frac{\dot{g} _m^{(A,iB)} }{g_m^{(A,iB)}}\right)^2 \Bigg] - m \right\},
 \ee
with $z=\tanh x$. Here dot denotes a derivative with respect to $z$.

According to the different conditions on the potential parameters they categorize the extended
potentials into three types. Out of these three, first two are isospectral to their conventional one.
Both of these extended potentials are equivalent, they differ only in the
range of the potential parameters.

Here we consider type I case and define the parameter $A^{'}=A+1$, and the other parameters as
\ba\label{am}
g_{m}^{(A,iB)}(z)&=&P_{m}^{(\alpha _{m},\beta _{m})}(z), \nonumber \\
\mbox{where}\qquad \alpha _{m}&=&A+1-m+\frac{iB}{A+1-m}, \nonumber\\
\beta _{m}&=&A+1-m-\frac{iB}{A+1-m};\qquad m=1,2,3,....; \quad A>m-1,
\ea
one obtains the rationally extended RM-II potentials, $V^{(-)}(x) =V_{(A,iB,ext)}(x)$ isospectral to the potentials
$V^{(+)}(x)$ with a bound state spectrum
\ba\label{extdrm2ev}
E_n=-(A +1-n )^2+\frac{B^2}{(A+1-n)^2}, \quad n=0,1,2....n_{max}, \qquad n_{max}<A+1.
\ea
The corresponding bound state eigenfunctions of $V^{(+)}(x)$ and $V^{(-)}(x)$ are given by
\ba\label{extdrm2wf}
\psi _{n}^{(+)}(x)\propto (1-z)^{\frac{\alpha _{n }}{2}}(1+z)^{\frac{\beta  _{n}}{2}}
P_{n}^{(\alpha _{n},\beta _{n})}(z) \nonumber \\
 \ea
 and
\ba\label{extdrm2wf2}
\psi ^{(-)}_{n}(x)\propto \hat{A}\psi^{(+)}_{n}(x)
\ea
respectively. Where the parameters $\alpha _{n}=A+1-n +\frac{iB}{A+1-n},\quad \beta _{n}=A+1-n -\frac{iB}{A+1-n}$ and operator
\ba\label{A}
\hat{A}=(1-z^{2})\frac{d}{dz}+\frac{iB}{A+1}+(A+1)z-\frac{2(m+\alpha _{m})(m+\beta _{m})}{2m+\alpha _{m}+\beta _{m}}
\frac{g_{m-1}^{(A-1,iB)}}{g_{m}^{(A,iB)}}.
\ea
Using $\hat{A}$ and $z$ in Eq. (\ref{extdrm2wf2}) and after  simple calculation the wavefunction $\psi ^{(-)}_{n}(x)$
 is given by
\ba\label{extdrm2wf3}
\psi ^{(-)}_{n }(x)&\propto &\frac{(1-z)^{\frac{\alpha _{n }}{2}}
(1+z)^{\frac{\beta _{n }}{2}}}{g_{m}^{(A,iB)}(z)} y_{\nu}^{(A,iB)}(z), \quad \nu = m+n -1
\ea
 where $y^{(A,iB)}_{\nu}(z)$ is some new type of $\nu$ th-degree polynomial in $z$ defined in terms of
classical Jacobi polynomials as
\ba\label{y}
y_{\nu}^{(A,iB)}(z)&=&\frac{2(n +\alpha _{n })(n +\beta _{n })}{2n +\alpha _{n }+\beta _{n }}g_{m}^{(A,iB)}(z)
P_{n-1}^{(\alpha _{n },\beta _{n})}(z)- \nonumber \\
&&\frac{2(m+\alpha _{m})(m +\beta _{m})}{2m+\alpha _{m}+\beta _{m}}g_{m-1}^{(A-1,iB)}(z)
P_{n}^{(\alpha _{n},\beta _{n})}(z).
\ea
The superpotential corresponding to this potential is given by
\ba
\bar{W}(x)&=&-\log(\psi^{(-)}_{0}(x))' \nonumber\\
&=& \frac{iB}{A+1}+(A+1)z-(1-z^2)\bigg( \frac{\dot{g}^{(A-1,iB)}_{m-1}}{g^{(A-1,iB)}_{m-1}}-\frac{\dot{g}^{(A,iB)}_{m}}{g^{(A,iB)}_{m}}\bigg ).
\ea
The partner potentials are 
\be
\bar{V}^{(\pm)}(x)=\bar{W}^2(x)\pm\bar{W}(x),
\ee
where 
\ba
\bar{V}^{(+)}(x)&=&V^{(-)}(x)=V^{(m)}_{(A,iB,ext)}(x) \nonumber\\
\mbox{ and} \qquad \bar{V}^{(-)}(x)&=&V^{(m-1)}_{(A-1,iB,ext)}(x).
\ea
Similar to the real extended case \cite{op4}, we see that
the above rationally extended complex $PT$ symmetric potential
 also satisfies an unfamiliar type of extended SI conditions where the partner potential $\bar{V}^{(-)}(x)$ is obtained by 
translating the potential parameter $A\rightarrow A-1$ (as in the usual SI condition) and the degree of the polynomial
$m\rightarrow m-1$.


\section{ Scattering amplitudes: REPTS complex potentials}

In the case of one dimensional systems, the transmission and reflection
amplitudes can be calculated easily by assuming the asymptotic behavior of
the wavefunctions at $x\longrightarrow \pm\infty$.

In this section, we first consider an asymptotically vanishing potential,
 the rationally extended complex Scarf II potential and then  a non-vanishing
 potential, the rationally extended
complex RM-II potential and obtain the scattering amplitudes in both the
cases.

\subsection{REPTS complex  Scarf II potential}

For simplicity, we first consider the potential whose bound state
eigenfunctions  are associated with the $X_1$ Jacobi polynomial
and then we generalize it to the $X_m$ case.
For $m=1$,  the bound state eigenfunctions of the rationally  extended complex
$PT$ symmetric Scarf II potential are given by Eq. (\ref{scarfextdwf}), i.e.
\be\label{psi1}
\psi_{n,1 }(x)\propto \frac{(1+\sinh^2 x)^{-\frac{A}{2}} \exp \left\{{-iB\tan^{-1}(\sinh x)}\right\}}{P^{(-\alpha-1 ,\beta-1 )}_{1} (i\sinh x)}
\hat{P}^{(\alpha ,\beta )}_{n +1} (i\sinh x),
\ee
where $\hat{P}^{(\alpha ,\beta )}_{n +1} (i\sinh x)$ is the $X_1$ Jacobi polynomial \cite{dnr1,dnr2} written in terms of classical Jacobi polynomial as
\ba\label{x1}
\hat{P}^{(\alpha ,\beta )}_{n +1} (i\sinh x) &=& \frac{1}{2(\alpha +\beta +2n )}\bigg[\left\{(b - i\sinh x )(\alpha +\beta + n )+ 2 b \right\}
P^{(\alpha ,\beta )}_{n} (i\sinh x)\nonumber \\
 &-&2 P^{(\alpha ,\beta )}_{n-1} (i\sinh x)\bigg],
\ea
with $b = \frac{\beta +\alpha }{\beta -\alpha }$.

Using the properties of Jacobi polynomials \cite{toi}
\ba\label{djacobi}
2P^{(\alpha ,\beta )}_{n-1} (z)=\frac{(2n +\alpha +\beta )(1-z^2)}{(n +\alpha )(n +\beta )}
\frac{d}{dz}P_{n }^{(\alpha ,\beta) }(z) -
\frac{n \big[ (\alpha -\beta )-(2n +\alpha +\beta )z\big]}
{(n +\alpha )(n +\beta )}P_{n }^{(\alpha ,\beta ) }(z), \nonumber\\
\ea
the $X_1$ Jacobi polynomial as given by Eq. (\ref{x1}) can be simplified in
terms  of $P_{n }^{(\alpha ,\beta ) }(i\sinh x)$
\ba\label{x11}
\hat{P}^{(\alpha ,\beta )}_{n +1} (i\sinh x) &=& \frac{1}{2(\alpha +\beta +2n )}\bigg[\bigg\{ \big( (b - i\sinh x )(\alpha +\beta + n )+ 2 b\big)\nonumber\\
&+ &\frac{n \big[ (\alpha -\beta )-(2n +\alpha +\beta )i\sinh x\big]}{(n +\alpha )(n +\beta )}\bigg\}P^{(\alpha ,\beta )}_{n} (i\sinh x)\nonumber \\
 &-&\frac{(2n +\alpha +\beta )}{(n +\alpha )(n +\beta )}\bigg(\frac{\cosh x}{i}\bigg)
\frac{d}{dx}P_{n }^{(\alpha ,\beta)}(i\sinh x)\bigg].
\ea
In terms of hypergeometric function, the classical Jacobi Polynomial $P^{(\alpha ,\beta )}_{n} (i\sinh x)$
 can be written as
\be\label{hyp}
P^{(\alpha ,\beta )}_{n} (i\sinh x) = (-1)^{n }\frac{\Gamma(n +\beta +1)}{n ! \Gamma(1+\beta )} F\left(n+\alpha +\beta +1, -n ;1+\beta
;\frac{1+i\sinh x}{2}\right).
\ee
Using Eqs. (\ref{x11}), (\ref{hyp}) and (\ref{psi1}), we get the bound state
eigenfunctions in terms of hypergeometric
functions.

To get the scattering state solutions, we must retain the second solution
which has been discarded in the case of bound states since it diverged
asymptotically. The second solution is included by replacing the
hypergeometric function $F(a,b,c,\tilde{z} )$ by
\be\label{hfun}
F(a,b;c;\tilde{z})=C_1F(a,b;c;\tilde{z})+C_2 {\tilde{z}}^{1-c}F(a-c+1, b-c+1;2-c;\tilde{z}),
\ee
where $C_1$ and $C_2$ are two constants. Further, instead of the parameter
$n$ labeling the number of nodes, one must use the
 wavenumber $k$ so that one gets the asymptotic wavefunctions  in terms of $e^{\pm ikx}$ as $x\rightarrow \pm\infty$.

To get the asymptotic wavefunctions at $x\rightarrow \pm\infty$ now we use another properties of Hypergeometric function \cite{toi}
\ba\label{hasym}
F(a ,b ;c ;\tilde{z}) &=& \frac{\Gamma(c )\Gamma(b -a )}{\Gamma(b)\Gamma(c-a )}
(-1)^a {\tilde{z}}^{-a} F(a ,a +1-c ;a +1-b ;\frac{1}{\tilde{z}})+ \nonumber \\
&&\frac{\Gamma(c )\Gamma(a-b  )}{\Gamma(a  )\Gamma(c -b )}
(-1)^b  {\tilde{z}}^{-b}  F(b  ,b  +1-c ;b  +1-a  ;\frac{1}{\tilde{z}}).
\ea

Now using Eq.(\ref{hasym}) and then taking the asymptotic behavior of the Eq. (\ref{psi1}) and after replacing $n$ by $A+ik$, we get
\ba\label{asym1}
\lim_{x\rightarrow {\infty }}\psi(k,x) &\simeq &\frac{exp(-\frac{i\pi B}{2})}{(-iB)}
\Bigg[\left(C_1 a_{k}+C_2 c_{k}(-1)^{(-\beta) }\right)
\left(\frac{1}{4i}\right)^{(A-ik)}\xi_1(k) e^{-ikx} \nonumber \\
&+&(C_1 b_{k}+C_2 d_{k}(-1)^{(-\beta)})\left(\frac{1}{4i}\right)^{(A+ik)}e^{ikx}\Bigg ].\nonumber
\ea
and
\ba
\lim_{x\rightarrow {-\infty }}\psi(k,x) &\simeq & \frac{exp(\frac{i\pi B}{2})}{(-i B)}\Bigg[\Big (C_1 a_{k}(-1)^{A-ik}+
C_2 c_{k}(-1)^{-(B+ik+\frac{1}{2})}\Big )\left(\frac{1}{4i}\right)^{A-ik}\nonumber\\
&\times &\xi_1(k)e^{ikx}+(C_1 b_{k}(-1)^{A+ik}+C_2 d_{k}(-1)^{(-B+ik-\frac{1}{2})})\nonumber \\
&\times &\left(\frac{1}{4i}\right)^{A+ik}e^{-ikx}\Bigg].
\ea
where
\ba
a_k =\frac{\Gamma(1+\beta )\Gamma(-2A-2ik -\alpha -\beta -1)}{\Gamma(-A-ik) \Gamma(-A-ik-\alpha )}, \ \
b_k =\frac{\Gamma(1+\beta )\Gamma(2A+2ik +\alpha +\beta +1)}{\Gamma(A+ik+\alpha +\beta +1) \Gamma(A+ik+\beta +1 )}, \nonumber
\ea

\ba
c_k =\frac{\Gamma(1-\beta )\Gamma(-2A-2ik -\alpha -\beta -1)}{\Gamma(-A-ik-\beta ) \Gamma(-A-ik -\alpha -\beta )}, \ \
\quad d_k =\frac{\Gamma(1-\beta )\Gamma(2A+2ik +\alpha +\beta +1)}{\Gamma(A+ik+\alpha +1) \Gamma(1+A+ik )} \nonumber
\ea
and
\be\label{xi1}
\xi_1(k) =  1+\frac{(2ik )}{(A+ik +\alpha )(A+ik +\beta )}.
\ee
The asymptotic behavior of the wave functions due to left incident  wave is given by
\ba\label{left1}
\lim_{x\rightarrow {-\infty }}\psi_{\nu }(x) &\simeq & e^{ikx} + r_{left}(k)e^{-ikx}\nonumber \\
\mbox{and}\qquad
\lim_{x\rightarrow {\infty }}\psi_{\nu }(x) &\simeq & t_{left}(k)e^{ikx} 
\ea

On comparing Eqs. (\ref{asym1}) and  (\ref{left1}), one can easily obtain
the constants $C_1$ and $C_2$

\be\label{C_1}
C_1=\frac{\exp(-\frac{i\pi B}{2})(\frac{1}{4i})^{(-A+ik)}(-1)^{(B+ik+\frac{1}{2})}}{a_k \{(-1)^{-\beta }-(-1)^{\beta }\}} \times \frac{1}{\xi_1(k)}
\ee
and
\be\label{C_2}
 C_2=\frac{-\exp(\frac{-i\pi B}{2})(\frac{1}{4i})^{(-A+ik)}(-1)^{(-A+ik)}}{c_k \{(-1)^{-\beta }-(-1)^{\beta }\}} \times \frac{1}{\xi_1(k)}.
\ee

In this way, we obtain the left transmission and reflection amplitudes

\be\label{tleft}
t_{left}(k)= t^{usual}_{left}(k)\times \left(\frac{(ik-\frac{1}{2})^2-B^2}{(ik+\frac{1}{2})^2-B^2}\right)\nonumber
\ee
and
\be\label{rleft}
r_{left}(k)= r^{usual}_{left}(k)\left(\frac{(ik-\frac{1}{2})^2-B^2}{(ik+\frac{1}{2})^2-B^2}\right),
\ee
where $t^{usual}_{left}(k)$ and $r^{usual}_{left}(k)$ are the transmission
and reflection amplitudes for the usual $PT$
symmetric complex Scarf II potential \cite{za} and are given by
\be\label{tleftu}
 t^{usual}_{left}(k)=\frac{\Gamma(-A-ik) \Gamma(1+A-ik) \Gamma(\frac{1}{2}-B-ik) \Gamma(\frac{1}{2}+B-ik)}{\Gamma(-ik)
\Gamma(1+ik) \Gamma^{2}(\frac{1}{2}-ik)},
\ee
and
\be\label{rleftu}
 r^{usual}_{left}(k)=t^{usual}_{left}(k)\times i\bigg[\frac{\cos \pi A \sin \pi B}{\cosh \pi k}+ \frac{\sin \pi A \cos \pi B}{\sinh \pi k}\bigg]
\ee
respectively.
Similar to the left incident wave, for the right incident case the asymptotic wavefunctions are
\ba\label{right1}
\lim_{x\rightarrow {\infty }}\psi_{\nu }(x) &\simeq & e^{-ikx} + r_{right}(k)e^{ikx}\nonumber \\
\mbox{and}\qquad
\lim_{x\rightarrow {-\infty }}\psi_{\nu }(x) &\simeq & t_{right}(k)e^{-ikx} 
\ea

By following similar procedure, for the case of the right incident wave, using Eqs. (\ref{asym1}) and (\ref{right1}) we obtain
\be\label{trightu}
t_{right}(k)=t_{left}(k)\nonumber
\ee
and
\be
r_{right}(k)=r^{usual}_{right}(k)\times \left(\frac{(ik-\frac{1}{2})^2-B^2}{(ik+\frac{1}{2})^2-B^2}\right),
\ee
with
\be
r^{usual}_{right}(k)=t^{usual}_{right}(k)\times i\big[-\frac{\cos(\pi A)\sin(\pi B)}{\cosh (\pi k)}+\frac{\cos(\pi B)\sin(\pi A)}{\sinh (\pi k)}\big ].
\ee

 {\bf\underline{Generalization to the $X_m$ case}:}


Generalization to the case of $X_m$ is straightforward.
We start from  Eq. (\ref{scarfextdwf}) where the
$n+m$ th degree $X_m$ Jacobi orthogonal polynomial
$\hat{P}^{(\alpha ,\beta )}_{n +m} (i\sinh x) $ can be expressed in terms of
the classical Jacobi polynomials by
 \ba\label{xmjacobi}
\hat{P}^{(\alpha ,\beta )}_{n +m} (i \sinh x) &=&\Bigg\{P^{(-\alpha-2,\beta  )}_{m}(i\sinh x)+
\frac{2n (m-\alpha +\beta -1)P^{(-\alpha,\beta)}_{m-1}(i\sinh x)}{(2m-\alpha +\beta -2)(2n +\alpha +\beta )} \nonumber \\
&-&\frac{n (\beta +m-1)P^{(-\alpha ,\beta )}_{m-2} (i \sinh x)}{(\alpha +n -m+1)(2m-\alpha +\beta -2)}\Bigg\}
P^{(\alpha ,\beta )}_{n}(i \sinh x)\nonumber \\
&+& \frac{(m-\alpha +\beta -1)(\alpha +n)}{(\alpha +n -m+1)(2n+\alpha +\beta )} P^{(-\alpha ,\beta )}_{m-1 }(i \sinh x)
P^{(\alpha ,\beta )}_{n -1 }(i \sinh x).\nonumber\\
\ea
On using Eq. (\ref{djacobi}) in the above equation
and by considering the normalization of the Jacobi polynomials
\be\label{normjacobi}
P^{(\alpha,\beta)}_m(i\sinh x)=\frac{\Gamma(\alpha+\beta+2m+1)}{ m! \Gamma(\alpha+\beta+m+1)}\Bigg(\frac{i\sinh x-1}{2}\Bigg)^m+
\mbox{(lower degree terms)},
\ee
and following the same steps as in the $X_1$ case
it is straightforward to obtain the scattering amplitudes $t_{left}(k,m)$
and $r_{left}(k,m)$  in the $X_m$ case. We obtain

\be\label{tleftxm}
t_{left}(k,m)= t^{usual}_{left}(k)\times \left(\frac{[B^2-(ik-\frac{1}{2})^2]+(B-ik+\frac{1}{2})(1-m)}
{[B^2-(ik+\frac{1}{2})^2]+(B+ik+\frac{1}{2})(1-m)}\right)\nonumber
\ee
and
\ba\label{rleftxm}
r_{left}(k,m)= r^{usual}_{left}(k)\times \left(\frac{[B^2-(ik-\frac{1}{2})^2]+(B-ik+\frac{1}{2})(1-m)}
{[B^2-(ik+\frac{1}{2})^2]+(B+ik+\frac{1}{2})(1-m)}\right).
\ea
Similarly, $t_{right}(k,m)$ and $r_{right}(k,m)$ for the $X_m$ case are
given by
\be\label{trightxm}
t_{right}(k,m)=t_{left}(k,m)\nonumber
\ee
and
\be
r_{right}(k,m)=r^{usual}_{right}(k)\times \left(\frac{[B^2-(ik-\frac{1}{2})^2]+(B-ik+\frac{1}{2})(1-m)}
{[B^2-(ik+\frac{1}{2})^2]+(B+ik+\frac{1}{2})(1-m)}\right).
\ee

As mentioned above, under the parametric symmetry
$B\leftrightarrow A+\frac{1}{2}$, while the conventional complex Scarf II
potential (\ref{scarfpot}) (and hence the corresponding scattering amplitudes)
are invariant, but the rationally extended complex Scarf II potential
(\ref{scarfextdpot}) is not invariant under this parametric transformation but
instead we get another extended complex Scarf II potential
Eq. (\ref{extdscarf1}).
  The scattering amplitudes of the new extended potential
Eq. (\ref{extdscarf1}) can be easily obtained by considering the asymptotic
behavior of the bound state solutions given in Eq. (\ref{extdscarfwf1}) or
by simply replacing the parameters $B\leftrightarrow A+\frac{1}{2}$
in Eqs. (\ref{tleftxm}) and (\ref{trightxm}).

 {\bf Remarks:}
 \begin{itemize}
 \item{It is interesting to note that as we go from the conventional to the rationally extended $X_m $case, we get the new expressions
for transmission and reflection amplitudes with an extra $m$ dependent term $\left(\frac{[B^2-(ik-\frac{1}{2})^2]+(B-ik+\frac{1}{2})(1-m)}
{[B^2-(ik+\frac{1}{2})^2]+(B+ik+\frac{1}{2})(1-m)}\right)$ multiplied with the transmission and reflection amplitudes
of the conventional $PT$ invariant potential.}
\item {We can check easily that for $m=0$ the above results correspond to the results of the conventional $PT$ symmetric Scarf II potential
\cite{za} and for $m=1$ the results reduce to the results of $X_1$ case as
given by Eqs. (\ref{tleftu}) and (\ref{trightu}). }
\item{As in the conventional case \cite{za}, even in the rationally extended
complex Scarf II case the reflection coefficient exhibits the handedness
effect i.e., $r_{left}(k)\ne r_{right}(k)$.  }
 \item{Since $\lvert\frac{[B^2-(ik-\frac{1}{2})^2]+(B-ik+\frac{1}{2})(1-m)}
{[B^2-(ik+\frac{1}{2})^2]+(B+ik+\frac{1}{2})(1-m)}\rvert=1$, hence $T(k)$
and $R(k)$ satisfy the same reciprocity,
unitarity and other scattering properties as satisfied by the convention
$PT$ symmetric Scarf II potential \cite{zafer2,za2}. }
 \end{itemize}

\subsection{REPTS complex RM-II potential}

This is an asymptotically nonvanishing potential. For such potentials
we define the wave numbers $k$ for $x<0$ and
$k'$ for $x>0$. The asymptotic wave functions in terms of $k$ and $k'$ are thus given by
\ba\label{left}
\lim_{x\rightarrow -\infty}\psi(x)&\simeq& e^{ik'x}+r_{left}(k,k')e^{-ikx}\nonumber\\
\mbox{and}\qquad
\lim_{x\rightarrow +\infty}\psi(x)&\simeq& t_{left}(k,k')e^{ik'x},
\ea
where $r_{left}(k,k')$ and $t_{left}(k,k')$ are reflection and transmission amplitudes due to left incident wave.
Similarly for the right incident wave we have
\ba\label{right}
\lim_{x\rightarrow \infty}\psi(x) & \simeq & e^{-ikx}+r_{right}(k,k')e^{ik'x}\nonumber\\
\mbox{and}\qquad \lim_{x\rightarrow -\infty}\psi(x) &\simeq & t_{right}(k,k')e^{-ikx}.
\ea
The reflectivity $R(k,k')$ and the transmitivity $T(k,k')$ are
\be
R(k,k')={\lvert{r(k,k')}\rvert}^2 \qquad \mbox{and}\qquad T(k,k')={\lvert{t(k,k')}\rvert}^2
\ee
respectively.

For this asymptotically non-vanishing potential, the bound state energy
eigenvalues and the eigenfunctions
are given by Eqs. (\ref{extdrm2ev}) and (\ref{extdrm2wf2}) respectively. For $B>0$, the non-vanishing part of
the potential will be emissive for $x>0$ and absorptive for $x<0$. Due to this reason the asymptotic wave numbers $k$ and $k'$
will be complex and expressed in terms of $\alpha_n$ and $\beta_n$ as
\ba
E_n-V^{(-)}(x\rightarrow +\infty) &=& E_n-2iB=-\alpha^2_n = k^2\nonumber\\
 \mbox{and}\qquad E_n-V^{(-)}(x\rightarrow -\infty) &=& E_n+2iB=-\beta^2_n = k'^2.
\ea
Thus the complex parameters $\alpha_n$ and $\beta_n$ can be written as
\be\label{kk'}
\alpha_n=ik' \qquad   \beta_n=ik.
\ee
Eq. (\ref{extdrm2wf}) when expressed in terms of hypergeometric function,
takes the form
\ba\label{vpwf}
\psi _{n}^{(+)}(x)\propto (1-\tanh x)^{\frac{\alpha _{n }}{2}}(1+\tanh x)^{\frac{\beta  _{n}}{2}}
F(n+\alpha_n+\beta_n+1, -n; 1+\alpha_n;\frac{1-\tanh x}{2}).
\ea
Now we also have to consider the second solution of the hypergeometric
function by using Eq. (\ref{hfun}).  The
wavefunction $\psi _{n}^{(+)}(x)$ at $x\longrightarrow \infty$ takes the form
\ba\label{vpwf2}
\lim_{x\rightarrow \infty}\psi _{n}^{(+)}(x)\propto \bigg( C_1e^{-\alpha_n x}+C_2e^{\alpha_n x}\bigg).
\ea
To get the asymptotic solutions at  $ x\rightarrow -\infty $ we use another
property of the hypergeometric functions \cite{toi}
\ba
F(a ,b ;c ;z)&=&\frac{\Gamma(c )\Gamma(c -a -b )}{\Gamma(c -a )\Gamma(c -b )}
F(a ,b ;a +b -c +1;1-z)+(1-z)^{c -a -b } \nonumber \\
&&\frac{\Gamma(c )\Gamma( a +b -c )}
{\Gamma( a )\Gamma( b )}F(c -a ,c -b ;c -a -b +1;1-z)
\ea
in Eq. (\ref{vpwf}) and get
\ba\label{vpwf3}
\lim_{x\rightarrow -\infty}\psi _{n}^{(+)}(x)\propto \bigg( (C_1\chi_1 + C_2\chi_2)e^{\beta_n x}+(C_1\chi_3+C_2\chi_4)e^{-\beta_n x}\bigg),
\ea
where
\ba
&&\chi_1=\frac{\Gamma(1+\alpha _\nu )\Gamma(-\beta _{\nu })}{\Gamma(-\nu -\beta _{\nu })\Gamma(1+\alpha _{\nu }+\nu )} ;\quad
\chi_3=\frac{\Gamma(1+\alpha _\nu )\Gamma(\beta _{\nu })}{\Gamma(\nu +\alpha _{\nu }+\beta _{\nu }+1)\Gamma(-\nu )} \nonumber \\
&& \chi_2=\frac{\Gamma(1-\alpha _{\nu })\Gamma(-\beta _{\nu })}{\Gamma(-\alpha _{\nu }-\beta _{\nu }-\nu )\Gamma(1+\nu )} ; \quad
\chi_4=\frac{\Gamma (1-\alpha _{\nu })\Gamma(\beta _{\nu }))}{\Gamma(\nu +\beta _{\nu }+1)\Gamma(-\nu -\alpha _{\nu })}.
\ea
Using Eq. (\ref{extdrm2wf}) the asymptotic wavefunctions of the rationally extended complex Rosen-Morse II potential
are given by
\ba\label{rmasym}
\lim_{x\rightarrow \infty}\psi _{\nu }^{-}(x)& \propto & \lim_{x\rightarrow \infty}\big[\hat{A} \psi _{\nu }^{+}(x)\big]\nonumber\\
&=& C_1(\alpha_m-ik')e^{-ik'x}+C_2(\alpha_m+ik')e^{ik'x},\nonumber
\ea
and
\ba
\lim_{x\rightarrow -\infty}\psi _{\nu }^{-}(x)& \propto & \lim_{x\rightarrow -\infty}\big[\hat{A} \psi _{\nu }^{+}(x)\big]\nonumber\\
&=& \big[(C_1\chi_1+C_2\chi_2 )(ik-\beta_m)\big]e^{ikx}+\big[(C_1\chi_3+C_2\chi_4 )(-ik-\beta_m)\big]e^{-ikx}.\nonumber\\
\ea
Now comparing Eq. (\ref{left}) with Eq. (\ref{rmasym}), we get the constants
\be
C_1=0; \qquad \mbox{and} \qquad C_2=\frac{1}{\chi_2 (ik-\beta_m)},
\ee
with the same $\alpha_m$ and $ \beta_m$ given in Eq. (\ref{am}).

Using $\chi_2$ and the above constants $C_1$ and $C_2$, we can easily obtain the scattering amplitudes due to left incident wave
\be\label{tlrm}
t_{left}(k,k',m)=\frac{(\alpha_m+ik' )}{(ik-\beta _{m})} \times t^{usual}_{left}(k,k'),\nonumber
\ee
and
\be\label{rlrm}
r_{left}(k,k',m)=\frac{(-\beta_m-ik )}{(-\beta _{m}+ik)} \times r^{usual}_{left}(k,k'),
\ee
where $t^{usual}_{left}(k,k')$ and  $r^{usual}_{left}(k,k')$ are the left transmission and reflection amplitudes for the
usual $PT$ symmetric complex Rosen-Morse potential \cite{lev} given by
\be
t^{usual}_{left}(k,k')=\frac{\Gamma(-A-1-\frac{ik^{'}}{2}-\frac{ik}{2})\Gamma(A+2-\frac{ik^{'}}{2}-\frac{ik}{2})}
{\Gamma(1-ik^{'})\Gamma(-ik)}
\ee
and
\be
r^{usual}_{left}(k,k')=\frac{\Gamma(ik)}{\Gamma(-ik)}\frac{\Gamma(-A-1-\frac{ik^{'}}{2}-\frac{ik}{2})
\Gamma(A+2-\frac{ik^{'}}{2}-\frac{ik}{2})}{\Gamma(-A-1-\frac{ik^{'}}{2}+\frac{ik}{2})\Gamma(A+2-\frac{ik^{'}}{2}+\frac{ik}{2})}.
\ee
Similarly from Eqs. (\ref{right}) and (\ref{rmasym}) we can easily obtain
\ba\label{trrm}
t_{right}(k,k',m)&=&\frac{(-\beta  _{m}-ik)}{(\alpha  _{m}-ik^{'})} \left(\frac{k^{'}}{k}\right) \frac{\Gamma(-A-1-\frac{ik^{'}}{2}-\frac{ik}{2})
\Gamma(A+2-\frac{ik^{'}}{2}-\frac{ik}{2})}{\Gamma(1-ik^{'})\Gamma(-ik)}\nonumber\\
&=&\left(\frac{\beta_m-ik}{\alpha_m-ik'}\right)\left(\frac{ik-\beta_m}{ik'+\alpha_m}\right)\left(\frac{k'}{k}\right)t_{left}(k,k',m)\nonumber
\ea
and
\be
r_{right}(k,k',m)=\frac{(\alpha  _{m}+ik^{'})}{(\alpha  _{m}-ik^{'})}\frac{\Gamma(ik^{'})}{\Gamma(-ik^{'})}\frac{\Gamma(-A-1-\frac{ik^{'}}{2}-\frac{ik}{2})
\Gamma(A+2-\frac{ik^{'}}{2}-\frac{ik}{2})}{\Gamma(-A-1+\frac{ik^{'}}{2}-\frac{ik}{2})\Gamma(A+2+\frac{ik^{'}}{2}-\frac{ik}{2})}.
\ee
 {\bf Remarks:}
 \begin{itemize}
\item { Similar to the extended $PT$ invariant complex Scarf II case, the
left/right scattering amplitudes for the extended complex Rosen-Morse II
potential are also modified with an extra $m$ dependent factor. As expected,
in the special case of $m=0$ these results match exactly with those of
complex $PT$ invariant RM-II.}
\item{Due to the asymptotically non-vanishing behavior of the extended
Rosen-Morse II potential, from Eqs. (\ref{tlrm}) and (\ref{trrm})
we observe that on changing the direction of the incoming wave, the
transmission amplitude too changes by a phase factor of
$\left(\frac{\beta_m-ik}{\alpha_m-ik'}\right)
\left(\frac{ik-\beta_m}{ik'+\alpha_m}\right)\left(\frac{k'}{k}\right)$.
Further, as in the extended complex Scarf II case,
the reflection amplitude for this potential also exhibits the handedness
effect.}
\item{Since the absolute value of the extra term is one, hence the scattering
 properties such as the unitarity and the reciprocity
will be the same for the complex extended $PT$ invariant as well as the
conventional complex $PT$ invariant Rosen-Morse II potential}

 \end{itemize}


\section{Summary and Conclusions}

In this paper, we have filled the missing gap by discussing the bound state
eigenfunctions of the rationally extended $PT$ symmetric
complex Scarf II  and Rosen-Morse II potentials. The eigenvalues of these
potentials are isospectral to their conventional counterparts and the
eigenfunctions are written in terms of $X_m$ EOPs and some new types of
polynomials respectively. Further, we have briefly discussed the
parametric symmetry related to the extended Scarf II potential.
The reflection and the transmission amplitudes have also been calculated
for these potentials and we have shown that as
we go from the case of the conventional to the extended case, we get the
modified scattering amplitudes which are now $m$ dependent.
Some of the important scattering properties related to these potentials such
as the handedness effect, the reciprocity and
the unitarity have also been discussed.

{\bf Acknowledgments}

BPM acknowledges the financial support from the Department of Science and Technology (DST), Govt. of India under SERC project sanction
grant No. SR/S2/HEP-0009/2012. AK would like to thank Indian National Science Academy (INSA) for the award of INSA senior 
scientist position at Savitribai Phule Pune University, Pune.

\end{document}